\documentclass[showpacs,preprintnumbers,amsmath,amssymb,floatfix]{revtex4}
\usepackage[german,english]{babel}
\usepackage{graphicx}
\usepackage{amsmath}
\usepackage{amsfonts}
\usepackage{amssymb}
\usepackage{epsfig}
\usepackage[hang,nooneline]{subfigure}
\newcommand{\insertplot}[5]{\begin{figure}
 \hfill\hbox to 0.05in{\vbox to #5in{\vfill
 \inputplot{#1}{#4}{#5}}\hfill}
 \hfill\vspace{-.1in}
 \caption{#2}\label{#3}
 \end{figure}}
\newcommand{\inputplot}[3]{
 \special{ps: plotfile #1}
}
 
\newcounter{fig}

\begin{document}

\title{\bf Wormholes Immersed in Rotating Matter}

\author{{\bf Christian Hoffmann $^{1,2}$}}
\email[{\it Email:}]{christian.hoffmann@uni-oldenburg.de}
\author{{\bf Theodora Ioannidou $^3$}}
\email[{\it Email:}]{ti3@auth.gr}
\author{{\bf Sarah Kahlen $^1$}}
\email[{\it Email:}]{sarah.kahlen@uni-oldenburg.de}
\author{{\bf Burkhard Kleihaus $^1$}}
\email[{\it Email:}]{b.kleihaus@uni-oldenburg.de}
\author{{\bf Jutta Kunz $^1$}}
\email[{\it Email:}]{jutta.kunz@uni-oldenburg.de}
\affiliation{
$^1$
Institut f\"ur Physik, Universit\"at Oldenburg, Postfach 2503,
D-26111 Oldenburg, Germany\\
$^2$
Department of Mathematics and Statistics, University of Massachusetts, Amherst,
Massachusetts, 01003-4525, USA\\
$^3$ Department of Mathematics, Physics and Computational Sciences, Faculty of Engineering,\\
Aristotle University of Thessaloniki
Thessaloniki, 54124,  Greece
}

\date{\today}

\begin{abstract}
We demonstrate that rotating matter sets the throat of an Ellis
wormhole into rotation, allowing for wormholes which possess
full reflection symmetry with respect to the two asymptotically 
flat spacetime regions. We analyze the properties of this new type 
of rotating wormholes and show that the wormhole geometry can change
from a single throat to a double throat configuration.
We further discuss the ergoregions and the lightring structure of 
these wormholes.
\end{abstract}

\maketitle

\section{Introduction}

In General Relativity the Ellis wormhole 
\cite{Ellis:1973yv,Bronnikov:1973fh,Kodama:1978dw,Ellis:1979bh,Morris:1988cz,Morris:1988tu,ArmendarizPicon:2002km,Sushkov:2005kj,Lobo:2005us,Lobo:2017oab}
connects two asymptotically flat spacetime regions by a throat. 
The non-trivial topology is
provided by the presence of a phantom field, a real scalar field with a
reversed sign in front of its kinetic term. 
Recently, the static Ellis wormhole has
been generalized to spacetimes with higher dimensions 
\cite{Torii:2013xba,Dzhunushaliev:2013jja}.
Moreover, rotating generalizations of the Ellis wormhole 
in four and five dimensions
have been found 
\cite{Dzhunushaliev:2013jja,Kleihaus:2014dla,Chew:2016epf,Kleihaus:2017kai}.

Since wormholes could be of potential relevance in astrophysics,
a number of both theoretical and observational studies have been
performed.
These include astrophysical searches for wormholes
\cite{Abe:2010ap,Toki:2011zu,Takahashi:2013jqa},
studies of their properties as gravitational lenses
\cite{Cramer:1994qj,Perlick:2003vg,Tsukamoto:2012xs},
studies of their shadows
\cite{Bambi:2013nla,Nedkova:2013msa}
or of the iron line profiles of thin accretion disks
surrounding them \cite{Zhou:2016koy}.

An Ellis wormhole carries no further fields besides the phantom field.
Here we are interested in the effect of matter fields on the wormhole.
To this end, we immerse the wormhole throat inside a
lump of matter. While this question has been considered before for
nuclear matter 
\cite{Dzhunushaliev:2013lna,Dzhunushaliev:2014mza,Aringazin:2014rva}
and bosonic matter 
\cite{Charalampidis:2013ixa,Dzhunushaliev:2014bya,Hoffmann:2017jfs},
the matter considered so far was non-rotating.

We here add a new twist to this quest by immersing the wormhole
throat inside rotating matter, which we take as composed of
a complex boson field,
since this allows for the possibility to impose
rotation on the bosonic field by the choice of an appropriate ansatz.
The rotation of the matter then implies 
a rotation of the spacetime, thus dragging the wormhole throat along.

As we will see, 
such a rotating wormhole spacetime can be fundamentally different
from a rotating wormhole without matter (except for the
phantom field). Namely, the rotating configuration can be fully symmetric
with respect to a reflection of the radial coordinate at its
throat or in the case of a double throat at its equator.
This does not hold for the rotating wormholes obtained previously
\cite{Dzhunushaliev:2013jja,Kleihaus:2014dla,Chew:2016epf,Kleihaus:2017kai},
where the rotation is enforced via a boundary condition in only
one of the two asymptotically flat regions.

In section II we discuss the theoretical setting
to obtain wormholes immersed in rotating bosonic matter
in four spacetime dimensions. 
Besides presenting the action,
this includes a discussion of the field equations,
the boundary conditions and the wormhole properties.
In section III we present the results of the numerical
calculations for this new class of rotating wormholes
and discuss their global charges, their geometric properties,
and their lightrings.
We conclude in section IV.

\section{Theoretical setting}

We consider General Relativity with a minimally coupled 
complex scalar field $\Phi$
and a phantom field $\Psi$ in four spacetime dimensions.
Besides the Einstein-Hilbert action
with curvature scalar $\cal R$ and coupling constant $\kappa=8\pi G$,
the action 
%
\begin{equation}
S=\int \left[ \frac{1}{2 \kappa}{\cal R} 
+{\cal  L}_{\rm M} \right] \sqrt{-g}\  d^4x  
 \label{action}
\end{equation}

contains the matter Lagrangian ${\cal  L}_{\rm M}$
\begin{equation}
{\cal  L}_{\rm M} =
\frac{1}{2}\partial_\mu \Psi\partial^\mu \Psi
 -\frac{1}{2} g^{\mu\nu}\left( \partial_\mu\Phi^* \partial_\nu\Phi
                            + \partial_\nu\Phi^* \partial_\mu\Phi 
 \right) - 
          m_{\rm b}^2 |\Phi|^2
\label{matter}
\end{equation}
where the kinetic term of the massless phantom field $\Psi$
carries the reverse sign as compared to the kinetic term of the
complex scalar field $\Phi$,
which possesses mass $m_{\rm b}$.

Variation of the action with respect to the metric
leads to the Einstein equations 
\begin{equation}
G_{\mu\nu}= {\cal R}_{\mu\nu}-\frac{1}{2}g_{\mu\nu}{\cal R} 
=  \kappa T_{\mu\nu}
\label{ee} 
\end{equation}
with stress-energy tensor
\begin{equation}
T_{\mu\nu} = g_{\mu\nu}{{\cal L}}_M
-2 \frac{\partial {{\cal L}}_M}{\partial g^{\mu\nu}} \ ,
\label{tmunu} 
\end{equation}
while variation with respect to the
scalar fields yields for the phantom field the equation
\begin{equation}
\nabla^\mu \nabla_\mu \Psi =0 
\label{epsi} 
\end{equation}
and for the complex scalar field
\begin{equation}
\nabla^\mu \nabla_\mu \Phi 
   = 
   m_{\rm b}^2 \Phi \ .
\label{ephi} 
\end{equation}

To allow for the non-trivial topology of the stationary spacetime,
we choose for the line element 
\begin{equation}
ds^2 = -e^{f} dt^2 
    +e^{q-f}\left[e^b(d\eta^2 + h d\theta^2)+ h \sin^2\theta
    (d\varphi -\omega dt)^2\right] \ ,
\label{lineel}
\end{equation}
where  $f$, $q$, $b$ and $\omega$ are functions of 
the radial coordinate $\eta$ and the polar angle $\theta$,
while $h = \eta^2 +\eta_0^2$ is an auxiliary function containing
the throat parameter $\eta_0$.
The coordinate $\eta$ then takes positive and negative 
values, i.e. $-\infty< \eta < \infty$, 
where the two limits $\eta\to \pm\infty$
correspond to two distinct asymptotically flat regions.

We parametrize the complex scalar field $\Phi$ via \cite{Schunck:1996}
\begin{equation}
\Phi(t,\eta,\theta, \varphi) 
  =  \phi (\eta,\theta) ~ e^{ i\omega_s t + i n \varphi} \ ,   \label{ansatzp}
\end{equation}
where $\phi (\eta,\theta)$ is a real function,
$\omega_s$ denotes the boson frequency, 
and the integer $n$ is the rotational quantum number,
which we employ to impose rotation on the field configuration ($n \ne 0$).
The phantom field $\Psi$ depends only on the coordinates $\eta$ and $\theta$,
\begin{equation}
\Psi(t,\eta,\theta, \varphi) 
=  \psi (\eta,\theta) \ .
\label{ansatzph}
\end{equation}

Substituting the above Ans\"atze into the Einstein equations 
and the matter field equations
leads to a set of six coupled partial differential equations
(PDEs) of second order for the unknown metric and matter functions.
Inspection of this system of PDEs shows that it
allows for reflection symmetric solutions 
with respect to $\eta \to - \eta$.

To solve this system of PDEs,
we have to impose an appropriate set of boundary conditions
at the boundaries of the domain of integration.
For the new class of reflection symmetric rotating wormholes,
which are asymptotically flat and globally regular,
we impose the following conditions:
(i) the metric and scalar field functions
$f,q,b,\omega,\phi$
vanish asymptotically for $\eta \to \pm \infty$,
(ii) the derivatives $\partial_\theta f,
\partial_\theta q,\partial_\theta \omega$ and the functions
$b,\phi$ vanish along the rotation axis, $\theta=0$,
(iii) the derivatives $\partial_\theta f,\partial_\theta q,
\partial_\theta b, \partial_\theta \omega, \partial_\theta \phi$
vanish in the equatorial plane, $\theta = \frac{\pi}{2}$.

For the phantom field the boundary conditions are more involved.
While regularity and reflection symmetry with respect 
to the equatorial plane also require the derivative
$\partial_\theta \Psi(\eta,\theta)$ to vanish at $\theta = 0$
and $\theta = \frac{\pi}{2}$,
the boundary condition for $\Psi$ 
can be chosen freely 
at one asymptotically flat boundary,
since only derivatives of the phantom field enter the PDEs, 
and is determined from the 
asymptotic form of the solutions at the other 
asymptotically flat boundary.
%

%
Once the solutions are obtained we can read off their physical properties.
Mass and angular momentum can be obtained from the asymptotic behaviour of 
the metric functions
\begin{equation}
f \longrightarrow \mp\frac{2 M_{\pm}}{\eta} \ , \ \ \ 
\omega \longrightarrow \frac{2 J_{\pm}}{\eta^3}  \ \ \ \ {\rm as} \ \eta \to \pm \infty \ .
\label{MJinfty}  
\end{equation}
The particle number $Q$ associated with the conserved current
of the complex scalar field is related to the angular momentum by 
\cite{Schunck:1996}
$$ J = n Q \ . $$

Of interest are also the geometrical properties of the solutions.
In order to determine the presence of throats and equators we consider the 
circumferential radius in the equatorial plane,
\begin{equation}
R_e(\eta) = \sqrt{h}\left. e^{\frac{q-f}{2}}\right|_{\theta = \pi/2} \ .
\label{Rade}  
\end{equation}
The minima of $R_e$ correspond to throats, whereas the local maxima correspond 
to equators. Since $R_e$ increases without bound in the asymptotic regions,
any equator resides between two throats.
The ergoregion of the solutions
is defined as the region where the time-time component of the
metric is positive, $g_{tt}>0$. 
Its boundary is referred to as the ergosurface.
Here $g_{tt}(\eta,\theta) = 0$.

The geodesic motion of massless particles in the equatorial plane
is determined from
\begin{eqnarray}
\dot{\eta}^2 & = &
\frac{L^2}{e^{(b-q)}}\left(\frac{E}{L} -V_+(\eta)\right)
                     \left(\frac{E}{L} -V_-(\eta)\right) \ ,
\label{doteta2} \\
& & {\rm with} \ \ V_\pm(\eta) = \left(\omega\mp\frac{e^{f-q/2}}{\sqrt{h}}\right) \ ,	     
\label{Vpm} 
\end{eqnarray}
where the derivative is with respect to some affine parameter.
For circular orbits $\dot{\eta}$ and $\frac{d\dot{\eta}}{d\eta}$ 
have to vanish.
Thus the extrema of the potentials $V_\pm(\eta)$ 
determine the location of the lightring, $\eta_{\rm lr}$, 
and the ratio $\frac{E}{L}$, i.e. $\frac{E}{L}=V_\pm(\eta_{\rm lr})$.

\section{Symmetric rotating wormhole solutions}

We have solved the above set of PDEs numerically, subject to the 
above boundary conditions, employing the package FIDISOL \cite{schonauer:1989},
a finite difference solver based on the Newton-Raphson method.
The problem has the following input parameters:
i) the boson frequency $\omega_s$ is varied in the interval 
$0 < \omega_s < m_b$
(note, that the boson mass is kept fixed thoughout, $m_b=1.1$),
ii) the throat parameter is mostly fixed to $\eta_0=1$,
though other values have also been employed
(note, that $\eta_0=0$ leads to topologically trivial boson stars,
where $\eta \ge 0$),
iii) for the rotational quantum number the lowest values are considered,
$n \in \{0,1,2\}$, where $n=0$ represents the non-rotating case.
In the following we present results
for $0.3 \leq \omega_s \leq  1.03$, since for values of $\omega_s$ 
outside this interval
the numerical errors are too large to be considered as reliable solutions
(in the rotating case).

\begin{figure}[t!]
\begin{center}
\mbox{\hspace{0.2cm}
\subfigure[][]{\hspace{-1.0cm}
\includegraphics[height=.25\textheight, angle =0]{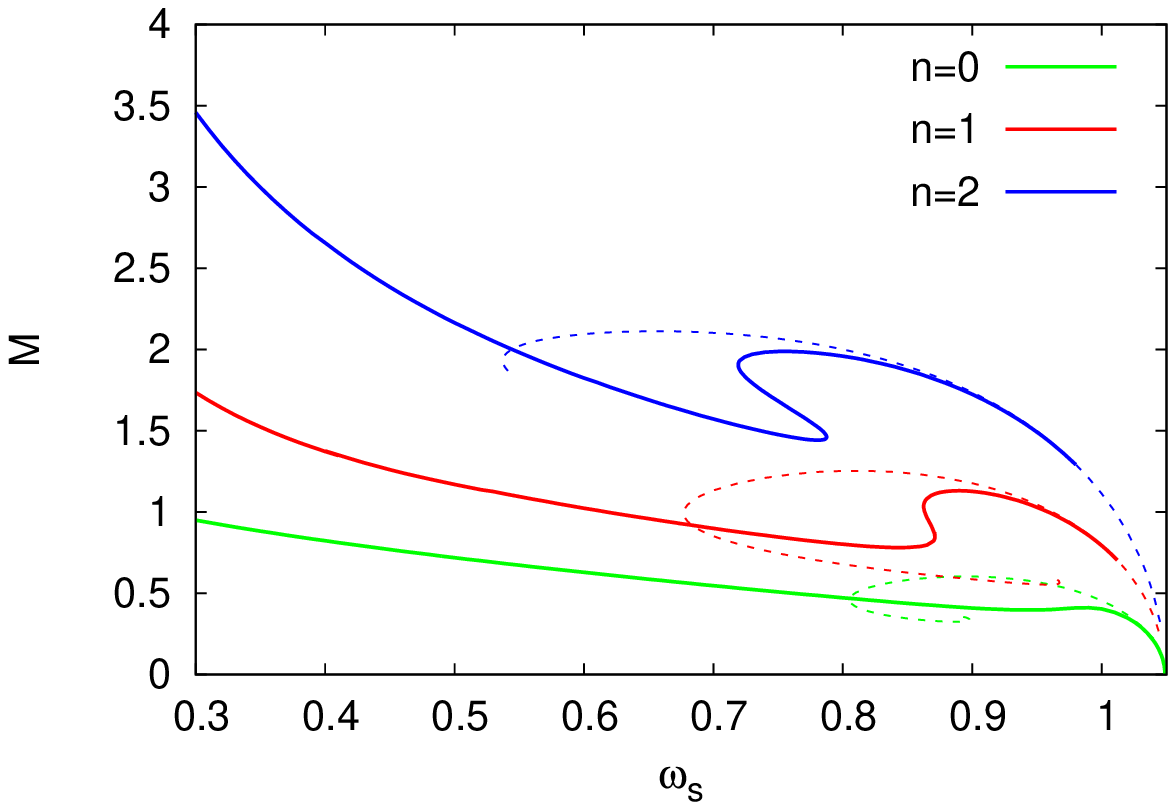}
\label{Fig1a}
}
\subfigure[][]{\hspace{-0.5cm}
\includegraphics[height=.25\textheight, angle =0]{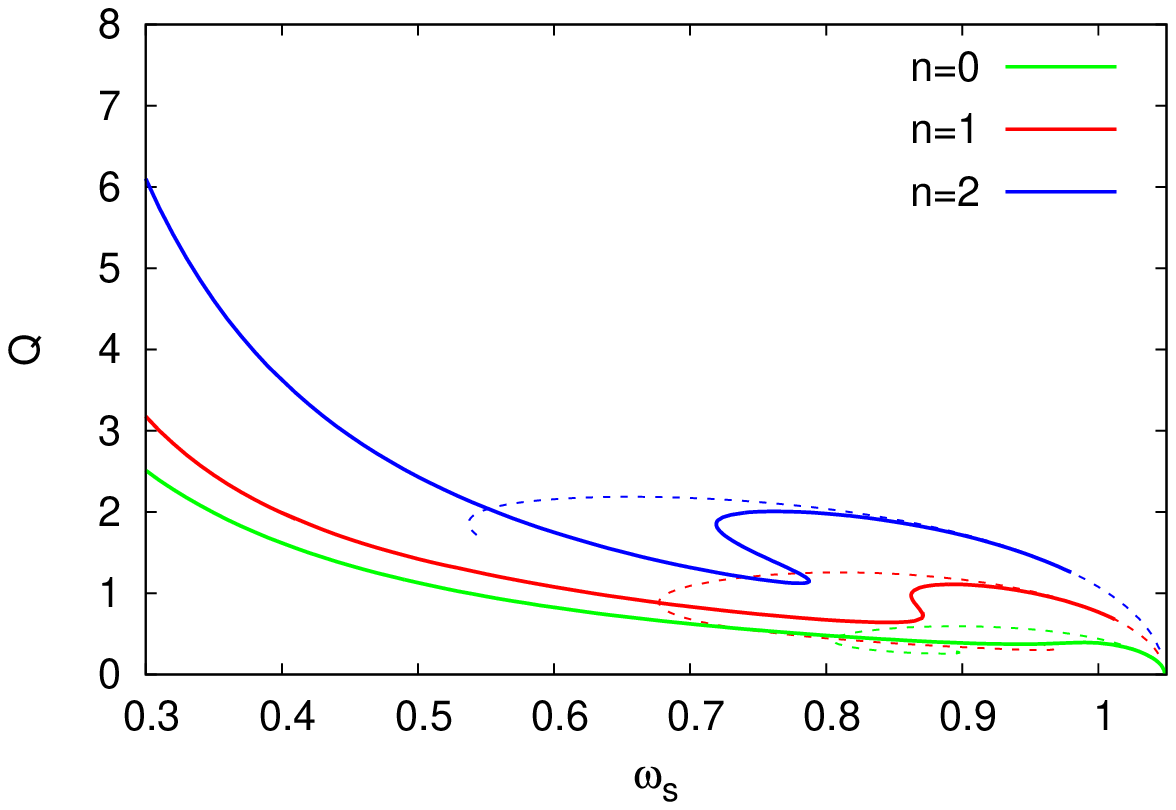}
\label{Fig1b}
}
}
\end{center}
\vspace{-0.5cm}
\caption{
Global charges of symmetric wormhole solutions
with throat parameter $\eta_0=1$
immersed in non-rotating ($n=0$) and rotating ($n=1$, 2) matter
versus the boson frequency $\omega_s$:
(a) the mass $M$; (b) the particle number $Q$.
These configurations possess angular momentum $J=nQ$.
The dashed lines indicate the corresponding boson star solutions.
\label{Fig1}
}
\end{figure}

Let us start our discussion of this new type of rotating wormhole solutions
by considering their global charges.
In Fig.\ref{Fig1} we show the mass $M$ and the particle number $Q$ 
versus the boson frequency $\omega_s$
for two families of rotating wormhole solutions,
specified by the lowest non-trivial rotational quantum numbers $n=1$ and $n=2$
and the throat parameter $\eta_0=1$.
The angular momentum $J$ of these solutions is given by $J=nQ$.
For comparison the figure also contains the 
corresponding set of non-rotating wormhole solutions 
(obtained for the same parameters except for $n=0$)
immersed in bosonic matter (see e.g.~\cite{Dzhunushaliev:2014bya}).

The domain of existence of these solutions is limited by
a maximal value for the boson frequency $\omega_s$,
which is reached for $\omega_{\rm max}=m_b$,
where the family of solutions reaches a vacuum configuration
with $M=0=Q$. 
This is completely analogous to the case of boson stars 
(see e.g.~\cite{Jetzer:1991jr,Lee:1991ax,Schunck:2003kk,Kleihaus:2005me,Kleihaus:2007vk,Liebling:2012fv}).
As seen in the figure, for large values of the frequency,
the global charges of the wormholes are similar to those of 
boson stars, which are also exhibited with thin black lines.
However the well-known spiralling behaviour of boson stars is largely lost.
Instead the would-be spirals unwind after their first backbending
with respect to the frequency and then continue to lower frequencies
(possibly all the way to $\omega_s=0$, where a singular
configuration should be reached \cite{Dzhunushaliev:2014bya}).
This unwinding is present in the case of rotating and non-rotating
solutions \cite{Dzhunushaliev:2014bya,Hoffmann:2017jfs},
and it has been seen as well in other systems with 
negative energy densities \cite{Hartmann:2013tca}.

\begin{figure}[t!]
\begin{center}
\mbox{\hspace{0.2cm}
\subfigure[][]{\hspace{-1.0cm}
\includegraphics[height=.25\textheight, angle =0]{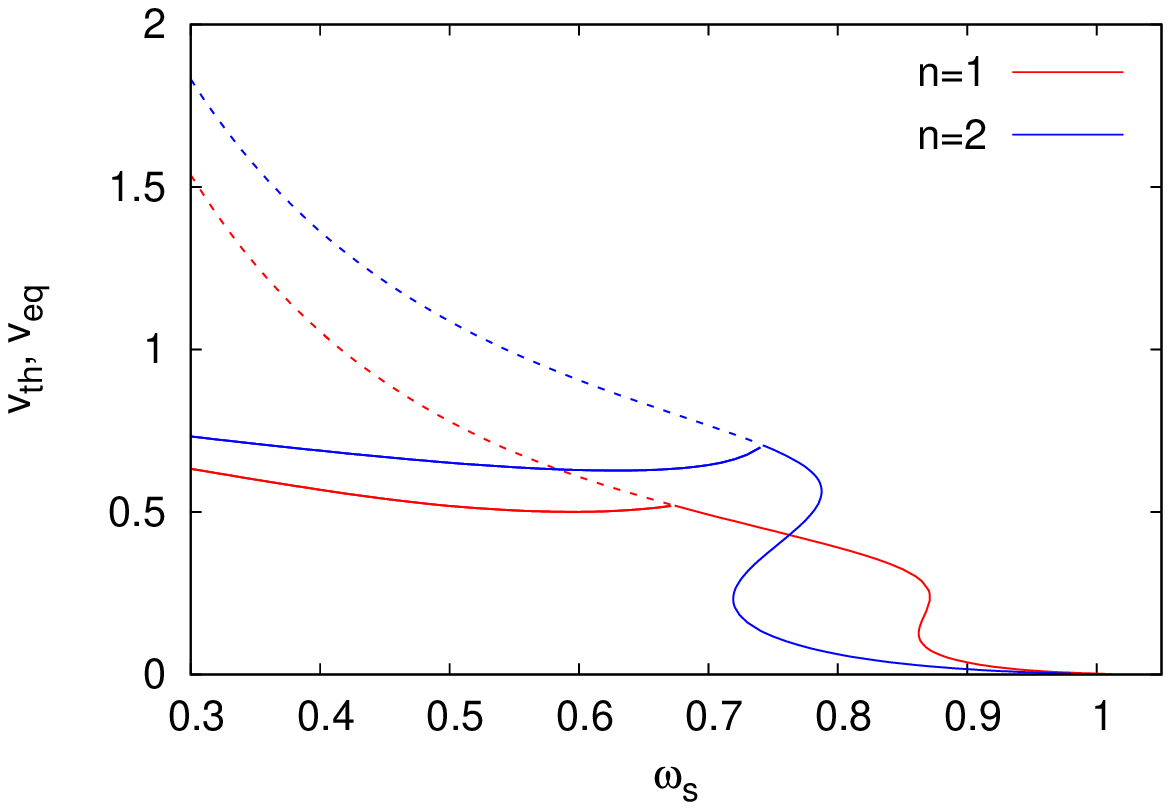}
\label{Fig2a}
}
\subfigure[][]{\hspace{-0.5cm}
\includegraphics[height=.25\textheight, angle =0]{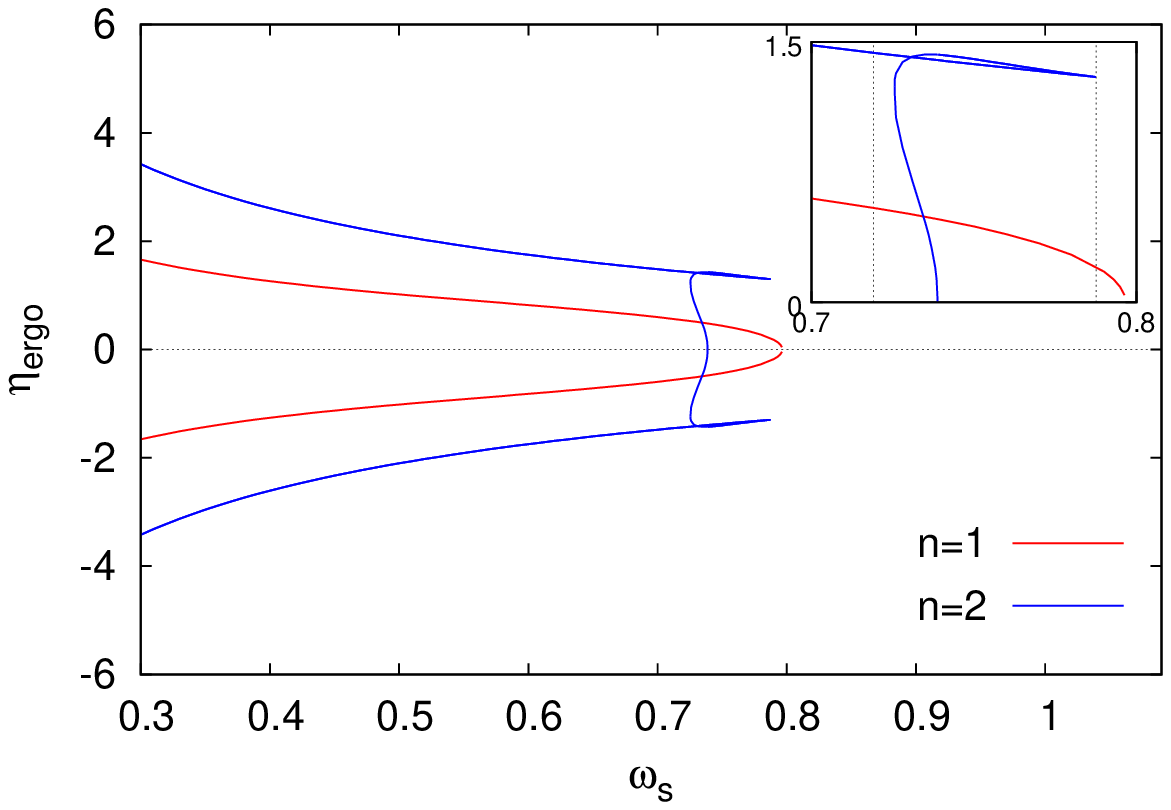}
\label{Fig2b}
}
}
\end{center}
\vspace{-0.5cm}
\caption{
Throat(s), equator and ergosphere(s) 
of symmetric wormhole solutions with throat parameter $\eta_0=1$
immersed in rotating matter
versus the boson frequency $\omega_s$:
(a) the rotational velocity $v_{\rm th}$ of the throat(s) 
(solid)
and the rotational velocity $v_{\rm eq}$ of the equator (dashed);
(b) the coordinate $\eta_{\rm ergo}$ of the ergosurface(s) 
in the equatorial plane.
\label{Fig2}
}
\end{figure}

Recall that we have imposed the same boundary conditions
in both asymptotic regions for the metric and the complex
scalar field. Thus we have not imposed rotation via
the boundary conditions. Instead, the ansatz for the
complex scalar field with a non-vanishing rotational 
quantum number $n$ imposes rotation on the configuration.

The rotation of the scalar field implies rotation of the
spacetime and consequently also rotation of the throat
of the wormholes. This rotation of the throat is demonstrated
for the above sets of rotating wormholes in Fig.~\ref{Fig2a},
where the rotational velocity $v_{\rm th}$
of the throat(s) in the equatorial plane 
is shown versus the boson frequency $\omega_s$.
Also shown is the rotational velocity $v_{\rm eq}$
of the equator when present.

We now turn to the ergoregions of the solutions.
In Fig.\ref{Fig2b} we show the coordinate $\eta_{\rm ergo}$
of the boundary of the ergoregions 
in the equatorial plane versus the boson frequency $\omega_s$. 
We note that ergoregions exist only if the boson frequency $\omega_s$
is smaller than some maximal value, which depends on $\eta_0$
and the rotational quantum number $n$.
For $n=1$ 
the ergoregion decreases monotonically with increasing $\omega_s$ 
and degenerates to a circle residing at $\eta_{\rm ergo}=0$,
when the maximal value of $\omega_s$ is approached. 

For $n=2$ the ergoregion decreases monotonically with increasing $\omega_s$ 
also up to a maximal frequency. 
But since this maximal frequency occurs in the range of the frequencies of the 
unwinding spiral, the presence of several branches of solutions
complicates the picture. 
Following the second branch (towards smaller $\omega_s$),
$\eta_{\rm ergo}$ reaches a maximum and then decreases again.
The ergoregion then disappears on the second branch 
before the third branch is reached.
Interestingly, for the small interval $0.7256 \leq \omega_s \leq 0.7388$ the
surface of the ergoregion consists of two disconnected parts.
At $\omega_s = 0.7388$, the throat at $\eta=0$ forms an inner boundary ring,
dividing the ergoregion into two parts, 
until at $\omega_s = 0.7256$ the ergoregion
degenerates to two rings. No ergoregion exists
for smaller frequencies on the second branch of solutions.

\begin{figure}[h!]
\begin{center}
\mbox{\hspace{0.2cm}
\subfigure[][]{\hspace{-1.0cm}
\includegraphics[height=.25\textheight, angle =0]{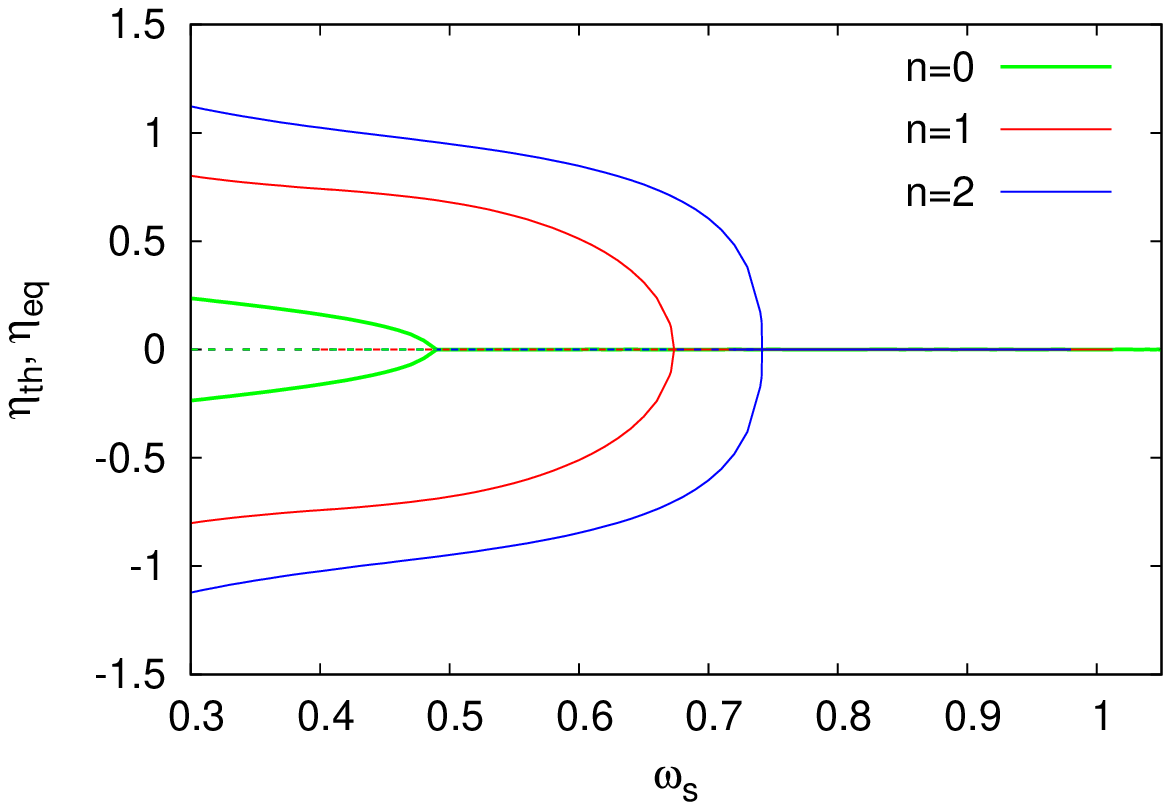}
\label{Fig3a}
}
\subfigure[][]{\hspace{-0.5cm}
\includegraphics[height=.25\textheight, angle =0]{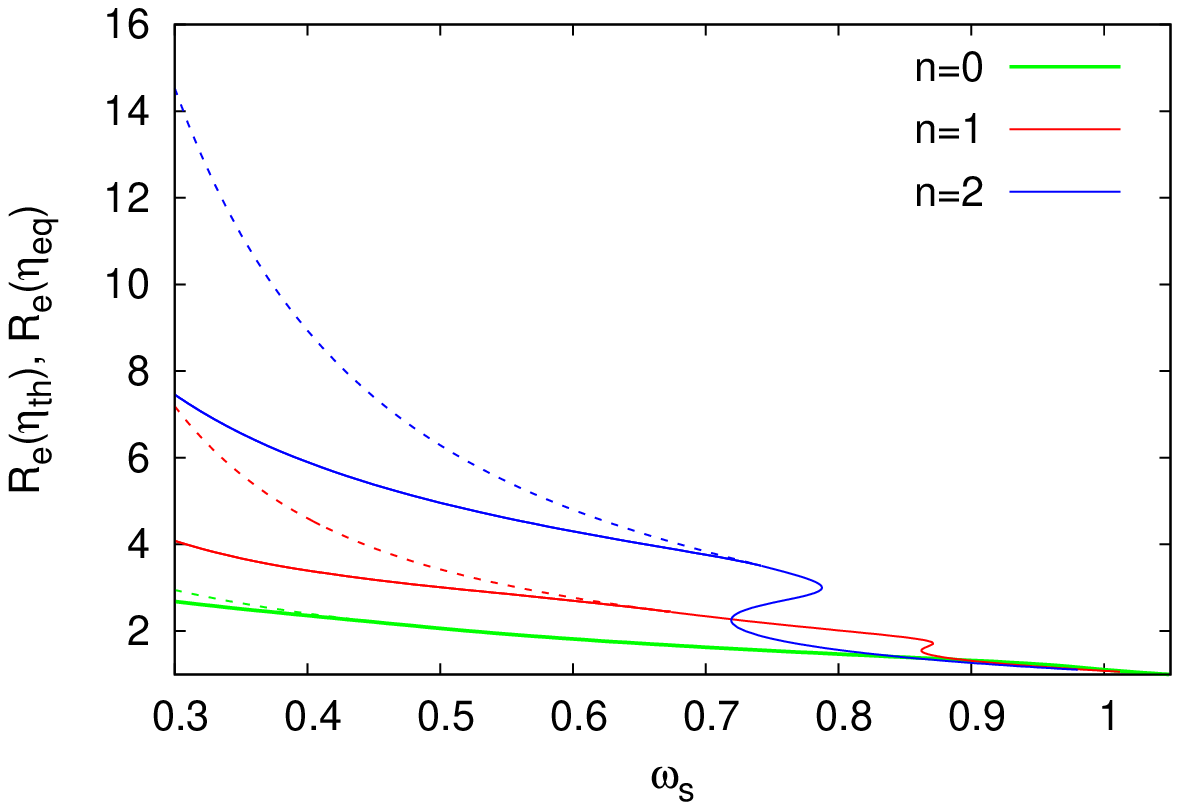}
\label{Fig3b}
}
}
\end{center}
\vspace{-0.5cm}
\caption{
Throat(s) and equator of symmetric wormhole solutions 
with throat parameter $\eta_0=1$
immersed in non-rotating ($n=0$) and rotating ($n=1$, 2) matter
versus the boson frequency $\omega_s$:
(a) the coordinate $\eta_{\rm th}$ of the throat(s) (solid)
resp.~the coordinate $\eta_{\rm eq}$ of the equator (dashed)
in the equatorial plane;
(b) the circumferential radius $R_{\rm e}(\eta_{\rm th})$
of the throat(s) (solid)
resp.~the circumferential radius $R_{\rm e}(\eta_{\rm eq})$
of the equator (dashed)
in the equatorial plane.
\label{Fig3}
}
\end{figure}

Let us next consider the geometry of the wormhole solutions.
In Fig.\ref{Fig3} we present the coordinate $\eta_{\rm th}$ of the throat(s) 
resp.~the coordinate $\eta_{\rm eq}$ of the equator in the equatorial plane
as well as the corresponding values of the circumferential radius $R_{\rm e}$.
We observe that for large values of the boson frequency $\omega_s$ 
the wormholes possess only a single throat. 
At a critical value of $\omega_s$ an equator emerges, 
when the throat degenerates to an 
inflection point at the critical value of $\omega_s$. 
For smaller $\omega_s$ the circumferential radius $R_{\rm e}$
possesses a maximum at $\eta=0$, corresponding to an equator, 
and two minima located symmetrically at around  $\eta=0$,  
corresponding to two throats. 

\begin{figure}[t!]
\begin{center}
\mbox{\hspace{0.2cm}
\subfigure[][]{\hspace{-1.0cm}
\includegraphics[height=.25\textheight, angle =0]{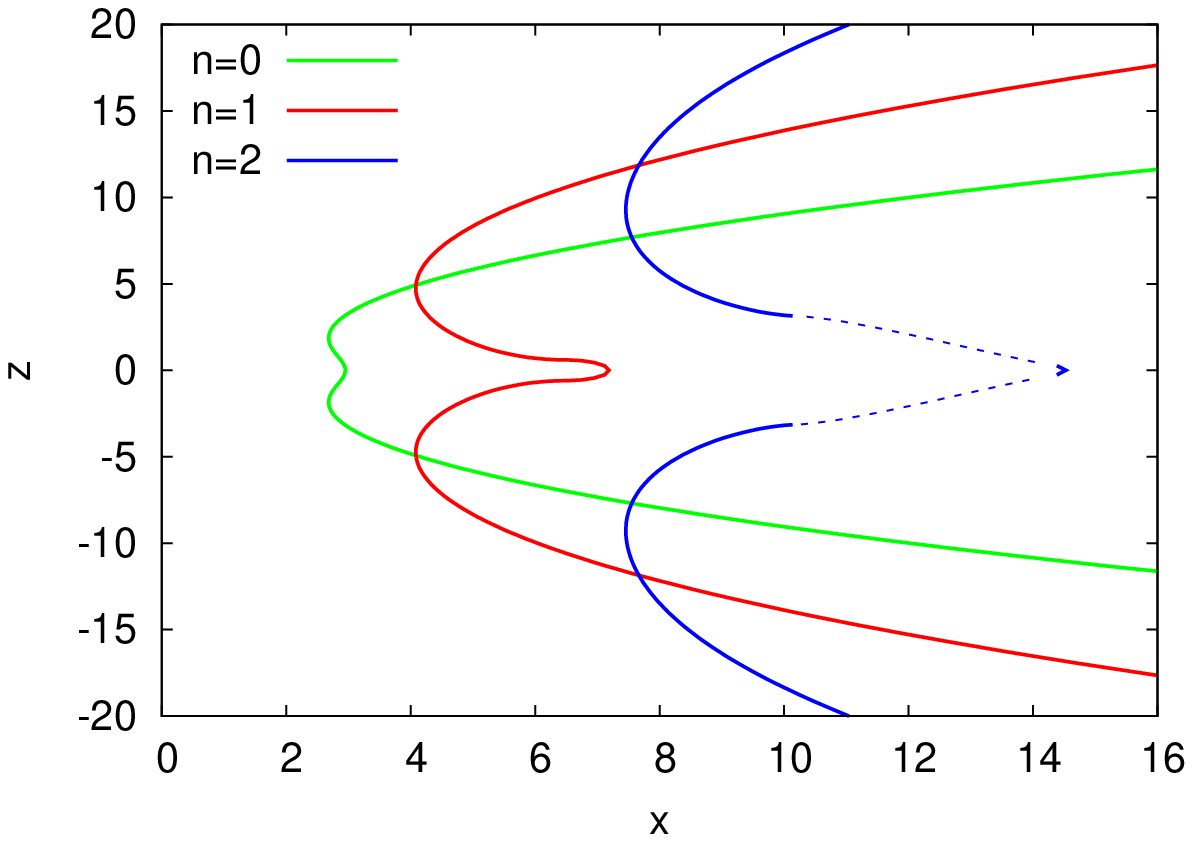}
\label{Fig4a}
}
\subfigure[][]{\hspace{-0.5cm}
\includegraphics[height=.25\textheight, angle =0]{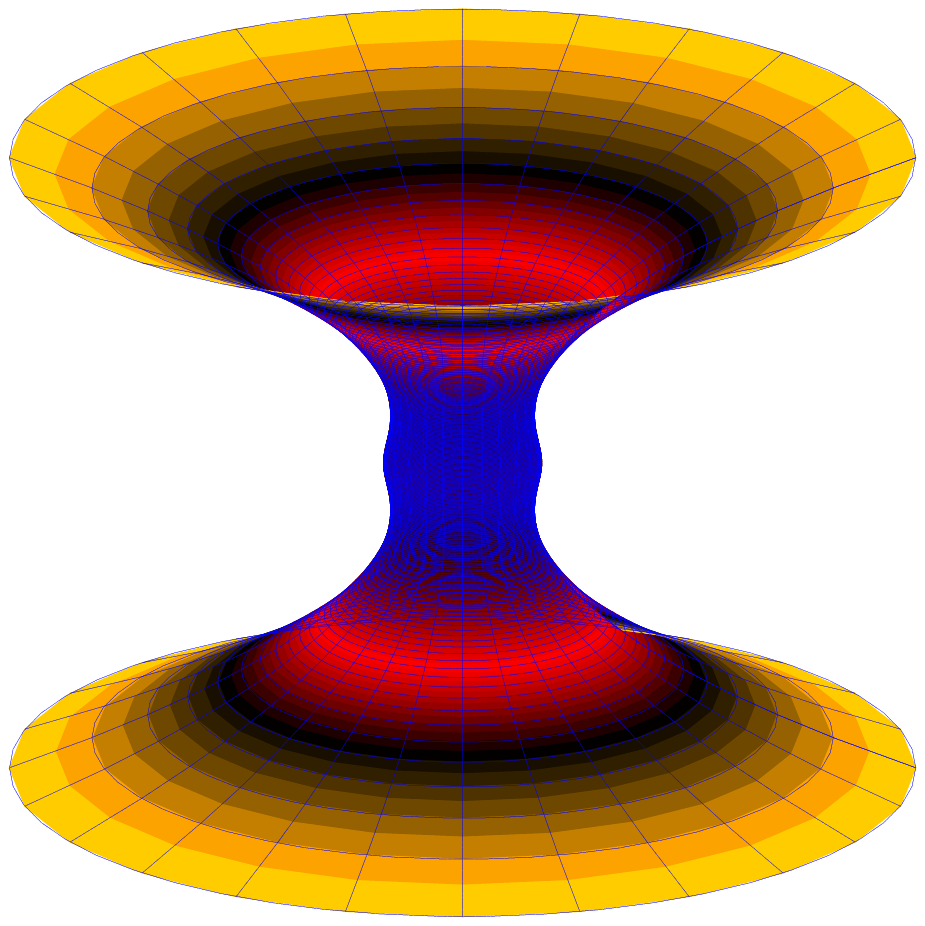}
\label{Fig4b}
}
}
\mbox{\hspace{0.2cm}
\subfigure[][]{\hspace{-1.0cm}
\includegraphics[height=.25\textheight, angle =0]{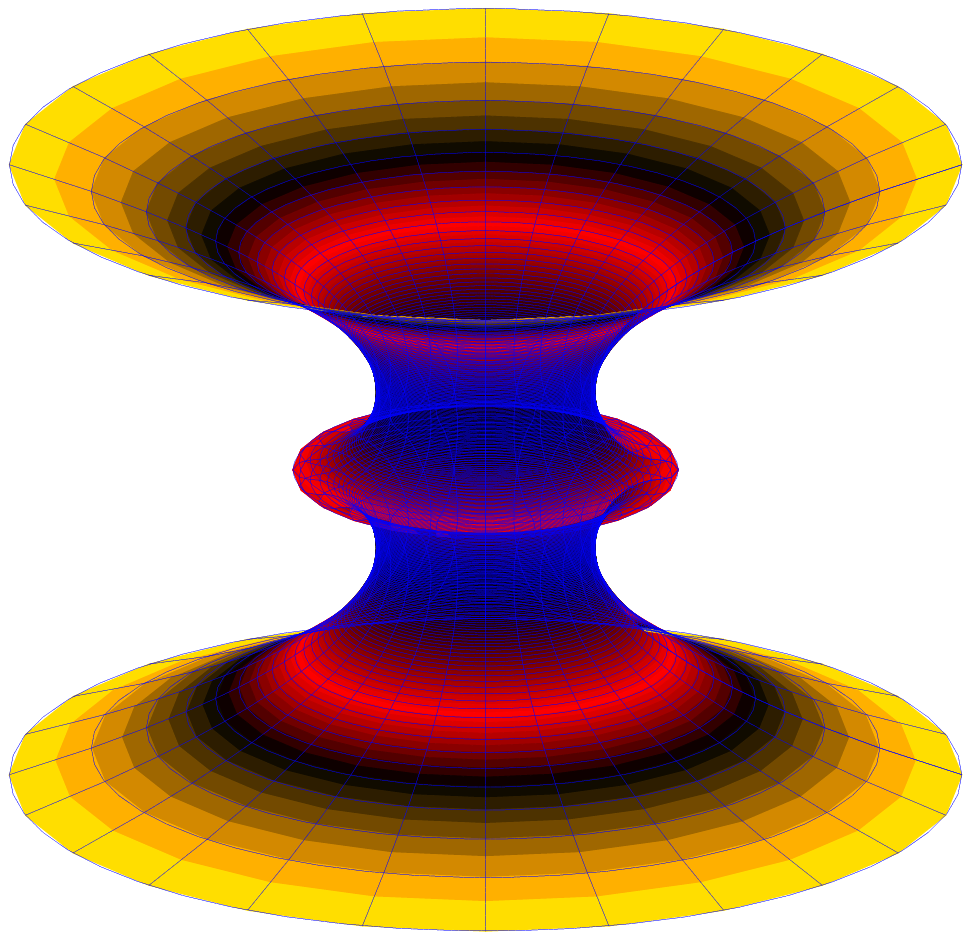}
\label{Fig4c}
}
\subfigure[][]{\hspace{-0.5cm}
\includegraphics[height=.25\textheight, angle =0]{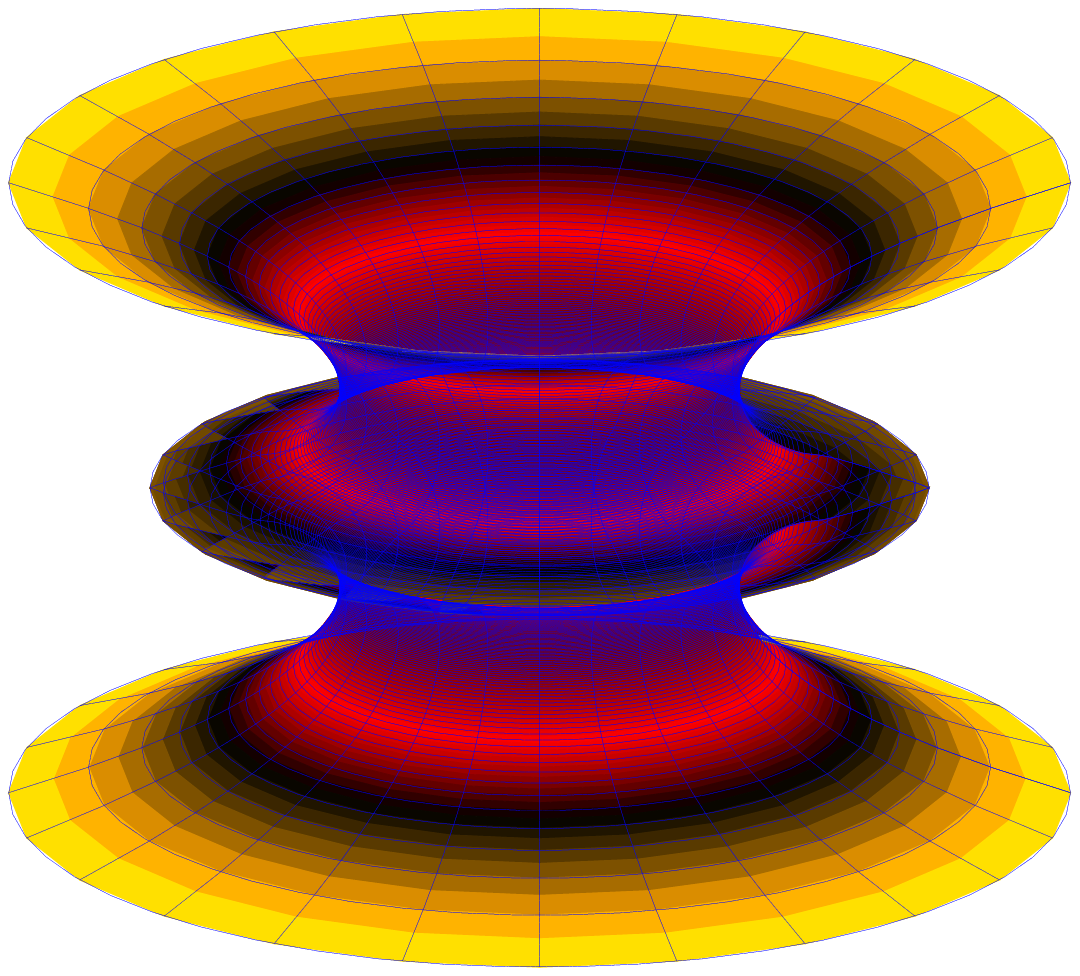}
\label{Fig4d}
}
}

\end{center}
\vspace{-0.5cm}
\caption{
Embeddings of the equatorial plane 
of symmetric wormhole solutions
with throat parameter $\eta_0=1$
immersed in non-rotating ($n=0$) and rotating ($n=1$, 2) matter:
(a) embeddings for $n=0$, 1, 2 and $\omega_s=0.3$,
where a pseudo-euclidean embedding is indicated by the dashed lines;
(b) $3D$ plot for $n=0$;
(c) $3D$ plot for $n=1$;
(d) $3D$ plot for $n=2$.
\label{Fig4}
}
\end{figure}

In order to get a deeper insight into the geometry of the wormholes,
we exhibit in Fig.\ref{Fig4} the isometric embedding of the equatorial plane 
for symmetric wormhole solutions immersed in non-rotating
($n=0$) and rotating ($n=1$, 2) matter
with $\eta_0=1$ and $\omega_s=0.3$
as examples of double-throat wormholes.
Note the pseudo-euclidean embedding for the case $n=2$.

\begin{figure}[t!]
\begin{center}
\mbox{\hspace{0.2cm}
\subfigure[][]{\hspace{-1.0cm}
\includegraphics[height=.25\textheight, angle =0]{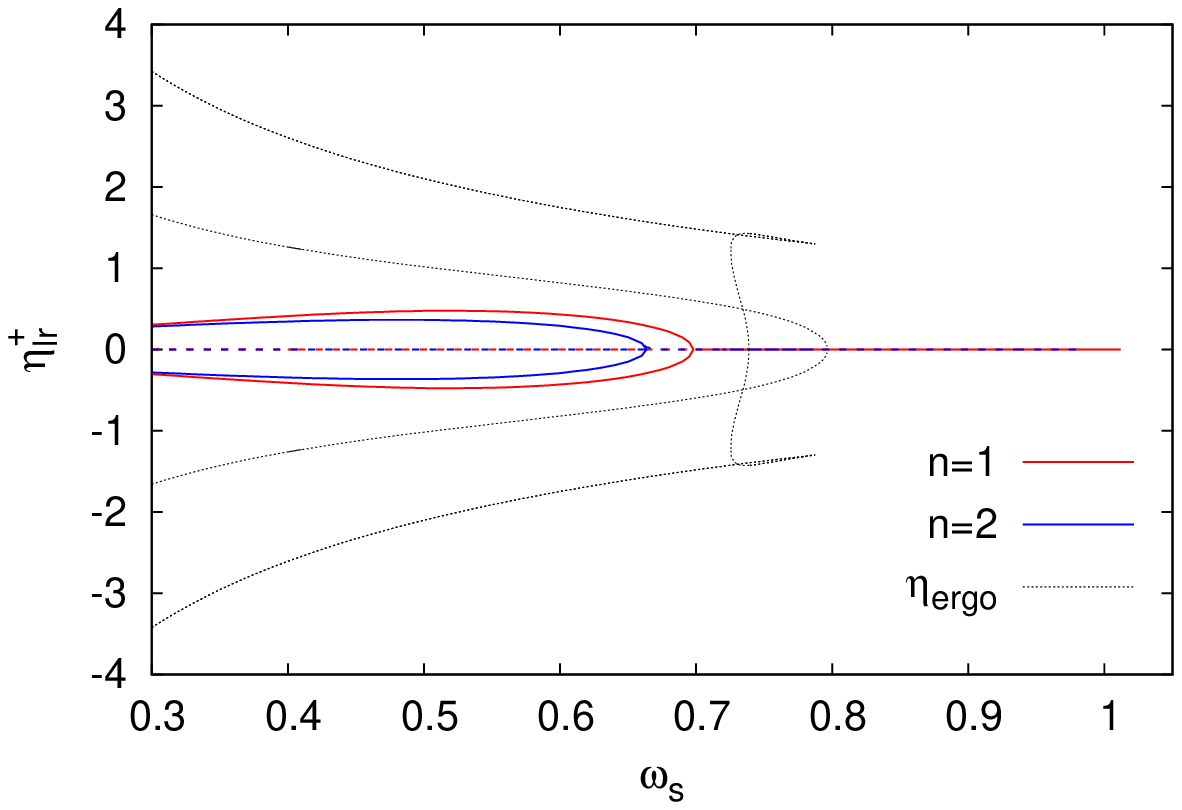}
\label{Fig5a}
}
\subfigure[][]{\hspace{-0.5cm}
\includegraphics[height=.25\textheight, angle =0]{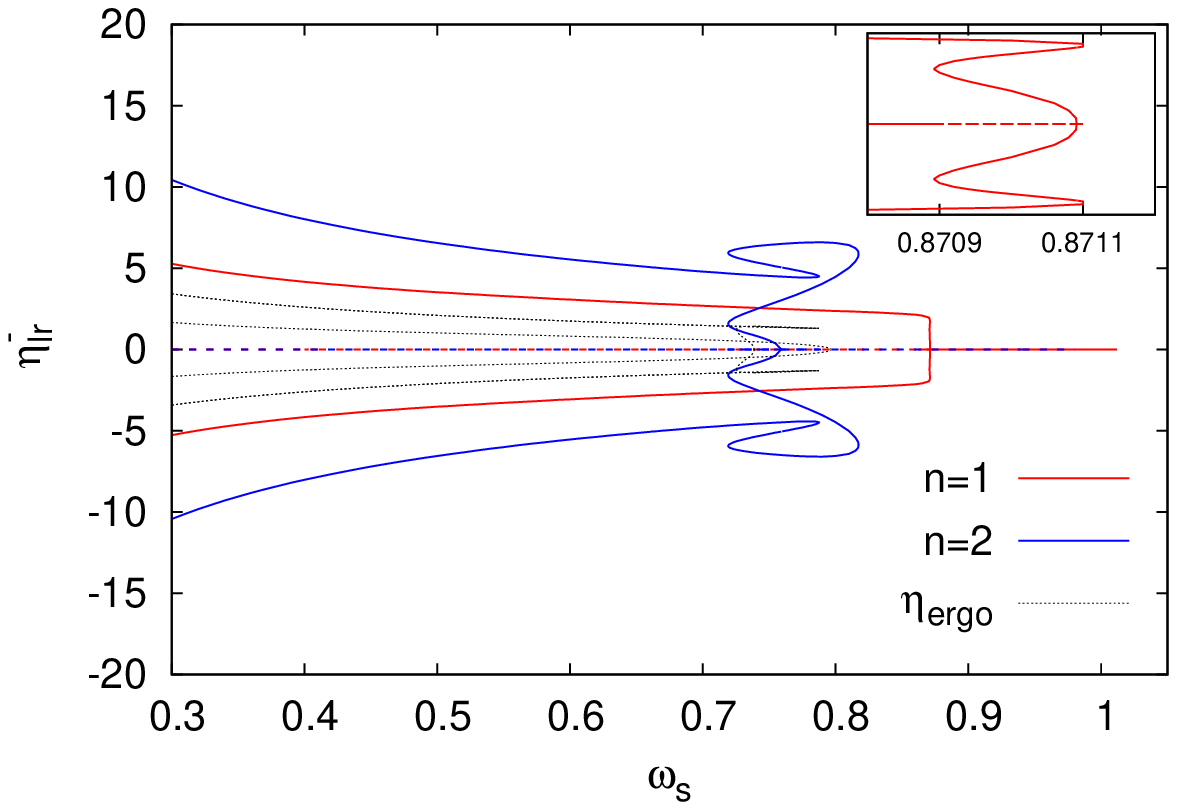}
\label{Fig5b}
}
}
\mbox{\hspace{0.2cm}
\subfigure[][]{\hspace{-1.0cm}
\includegraphics[height=.25\textheight, angle =0]{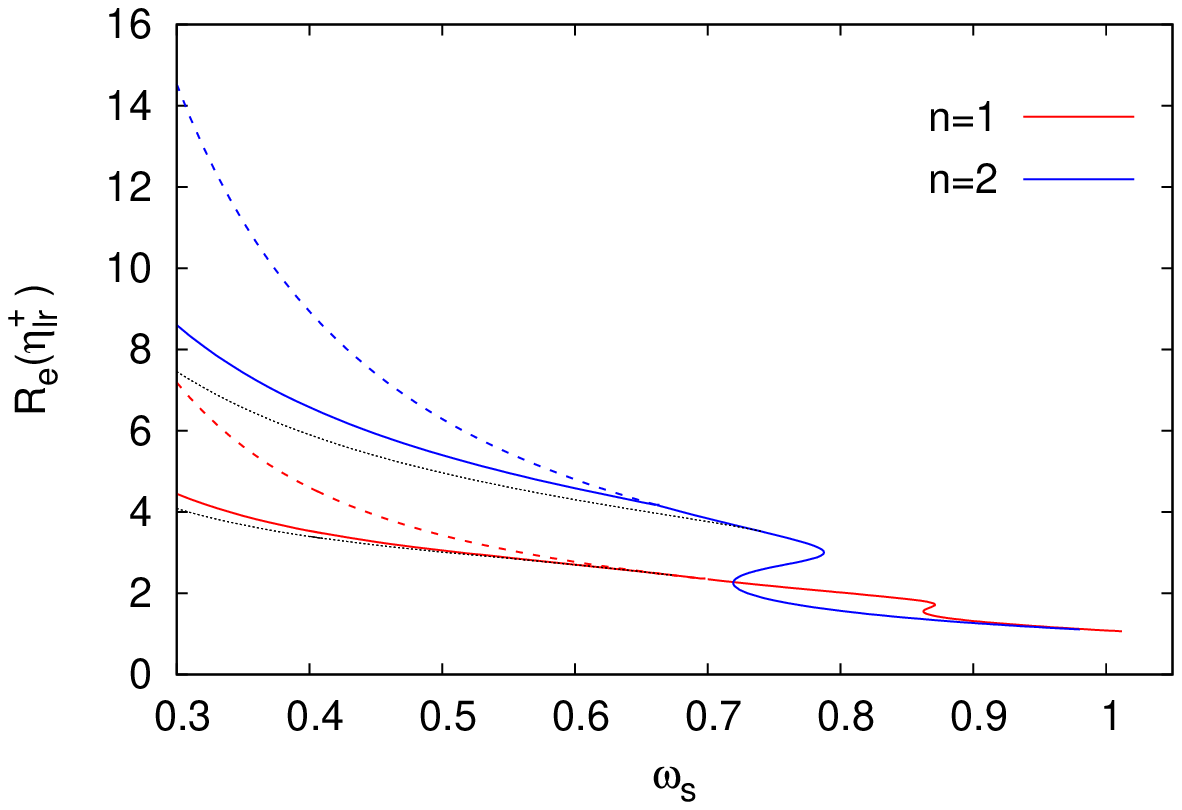}
\label{Fig5c}
}
\subfigure[][]{\hspace{-0.5cm}
\includegraphics[height=.25\textheight, angle =0]{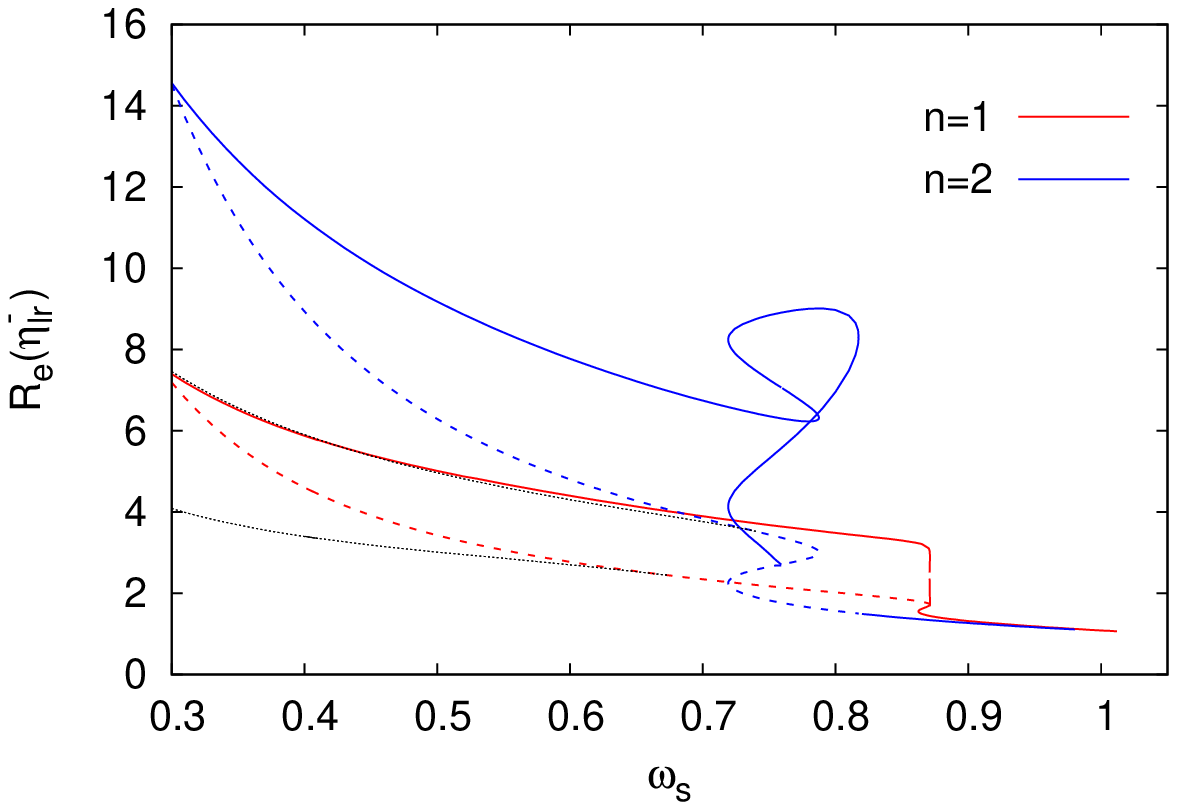}
\label{Fig5d}
}
}
\end{center}
\vspace{-0.5cm}
\caption{
Lightrings of symmetric wormhole solutions
with throat parameter $\eta_0=1$
immersed in rotating ($n=1$, 2) matter
versus the boson frequency $\omega_s$:
(a) the coordinate $\eta^+_{\rm lr}$ of co-rotating
massless particles in the equatorial plane;
(b) the coordinate $\eta^-_{\rm lr}$ of counter-rotating
massless particles in the equatorial plane;
(c) the circumferential radius $R_{\rm e}(\eta^+_{\rm lr})$;
(d) the circumferential radius $R_{\rm e}(\eta^-_{\rm lr})$.
Also shown are the coordinates and the circumferential radii
of the ergosurfaces (black lines).
\label{Fig5}
}
\end{figure}

Finally we would like to address the lightrings of these spacetimes.
In Fig.\ref{Fig5} we therefore exhibit the coordinate $\eta^+_{\rm lr}$ 
of co-rotating massless particles in the equatorial plane together with the
coordinate $\eta^-_{\rm lr}$ of counter-rotating massless particles
and their circumferential radii $R_{\rm e}(\eta^+_{\rm lr})$
and $R_{\rm e}(\eta^-_{\rm lr})$, respectively.
Also shown is the location of the ergosurfaces
and their respective circumferential radii (solid black).

The general picture is that a single lightring exists, if $\omega_s$ exceeds
a critical value, which depends on the orientation of the orbit. 
When  $\omega_s$ is smaller than this critical value 
two more lightrings emerge.
For the symmetric wormholes one of the lightrings 
is always located at $\eta=0$, 
i.e., at the throat, respectively, at the equator, 
whereas the remaining ones (if present) 
are located symmetrically around $\eta=0$.
We note that for symmetric wormhole solutions 
immersed in rotating ($n=2$) matter
up to five lightrings of counter-rotating massless
particles can exist.

When inspecting the circumferential radii for co-rotating and
counter-rotating massless particles, 
which are also exhibited in Fig.\ref{Fig5},
we realize that the circumferential radius $R_{\rm e}(\eta^+_{\rm lr})$
of the lightrings of co-rotating massless particles 
and the circumferential radius $R_{\rm e}(\eta_{\rm th})$
of the single throat
almost coincide for $n=1$ and are still close for $n=2$.

We further remark that the ratio $E/L$ of the lightring residing at
the equator of the symmetric wormholes becomes negative, 
when $\omega_s$ is smaller than a critical value. 
The change of sign occurs precisely, when the lightring
enters the ergoregion.

\section{Conclusion}

Whereas previously 
the rotation of a wormhole spacetime had to be imposed by hand,
by requiring a set of non-identical asymptotic boundary conditions
for the metric in the two universes
\cite{Kleihaus:2014dla,Chew:2016epf},
we have here obtained a new type of rotating wormhole solutions,
which possess identical asymptotic boundary conditions for the metric.
Indeed, instead of imposing the rotation as a boundary condition
we have obtained fully symmetric configurations
by immersing the wormholes into rotating matter.

For the rotating matter we have chosen for simplicity
a complex massive scalar field without any self-interaction.
Both asymptotically flat regions - or universes - possess the same global charges,
consisting of the mass, the angular momentum and the particle number,
where the latter two are related by the rotational quantum number
as in the case of boson stars.
We have shown that the resulting configurations
can possess interesting physical properties.
They can have a single throat or, depending on the parameters,
evolve an equator surrounded by a double throat.
These rotating wormhole solutions may also exhibit up to five
lightrings for counter-rotating massless particles.

The inclusion of a phantom field has allowed for the existence
of these topologically non-trivial solutions.
Clearly, by replacing General Relativity by certain generalized
gravity theories the phantom field would no longer be needed
\cite{Hochberg:1990is,Fukutaka:1989zb,Ghoroku:1992tz,Furey:2004rq,Bronnikov:2009az,Kanti:2011jz,Kanti:2011yv,Lobo:2009ip,Harko:2013yb}.
A study of rotating wormholes in such generalized theories
would therefore be desirable.

Let us end with a short remark concerning the stability
of the above spacetimes. It is known, that
the static Ellis wormhole in $D$ dimensions possesses an unstable radial mode
\cite{Shinkai:2002gv,Gonzalez:2008wd,Gonzalez:2008xk,Torii:2013xba}.
Interestingly, when the wormhole throat rotates sufficiently fast,
this radial instability disappears, 
as shown for (cohomogeneity-1) wormholes in five dimensions
\cite{Dzhunushaliev:2013jja}:
As the throat is set into rotation, a non-normalizable zero mode
of the Ellis wormhole turns into a second unstable radial mode.
With increasing rotation velocity of the throat, 
the eigenvalue of the original negative mode decreases, 
while the eigenvalue of the second unstable mode increases, 
until both merge and disappear.
Thus rotation has a stabilizing effect on the wormhole.
A stability analysis along these lines for
rotating 4-dimensional wormholes represents a challenging future task.

\section{Acknowledgments}

We would like to acknowledge support by the DFG Research Training Group 1620
{\sl Models of Gravity} as well as by FP7, Marie Curie Actions, People,
International Research Staff Exchange Scheme (IRSES-606096),
COST Action CA16104 {\sl GWverse}.
BK gratefully acknowledges support
from Fundamental Research in Natural Sciences
by the Ministry of Education and Science of Kazakhstan.

\end{document}